\documentclass[%
 aip,
 amsmath,amssymb,
 reprint,%
]{revtex4-1}
\usepackage{comment}
\usepackage{graphicx}
\usepackage{dcolumn}
\usepackage{bm}

\usepackage[utf8]{inputenc}
\usepackage[T1]{fontenc}
\usepackage{mathptmx}
\usepackage{etoolbox}
\usepackage{xcolor}

\newcommand{\un}[1]{\ensuremath{\mathrm{\,#1}}} 

\makeatletter
\def\@email#1#2{%
 \endgroup
 \patchcmd{\titleblock@produce}
  {\frontmatter@RRAPformat}
  {\frontmatter@RRAPformat{\produce@RRAP{*#1\href{mailto:#2}{#2}}}\frontmatter@RRAPformat}
  {}{}
}%
\makeatother
\begin{document}

\preprint{AIP/123-QED}

\title[Detection Efficiency Characterization for Free-Space SPDs: Measurement Facility and Wavelength-Dependence Investigation]{Detection Efficiency Characterization for Free-Space Single-Photon Detectors: Measurement Facility and Wavelength-Dependence Investigation}
\author{S. Virz\`i}
\email{s.virzi@inrim.it}
\affiliation{Istituto Nazionale di Ricerca Metrologica, str. delle Cacce 91, 10135 Turin, Italy}

\author{A. Meda}
\email{a.meda@inrim.it}
\affiliation{Istituto Nazionale di Ricerca Metrologica, str. delle Cacce 91, 10135 Turin, Italy}
\author{E. Redolfi}
\affiliation{Istituto Nazionale di Ricerca Metrologica, str. delle Cacce 91, 10135 Turin, Italy}
\affiliation{Physis Department, University of Turin, via P. Giuria 1, 10125 Turin, Italy}
\affiliation{Istituto Nazionale di Fisica Nucleare, sez. di Torino, via P. Giuria 1, 10125 Turin, Italy}
\author{M. Gramegna}
\affiliation{Istituto Nazionale di Ricerca Metrologica, str. delle Cacce 91, 10135 Turin, Italy}
\author{G. Brida}
\affiliation{Istituto Nazionale di Ricerca Metrologica, str. delle Cacce 91, 10135 Turin, Italy}
\author{M. Genovese}
\affiliation{Istituto Nazionale di Ricerca Metrologica, str. delle Cacce 91, 10135 Turin, Italy}
\affiliation{Istituto Nazionale di Fisica Nucleare, sez. di Torino, via P. Giuria 1, 10125 Turin, Italy}
\author{I. P. Degiovanni}
\affiliation{Istituto Nazionale di Ricerca Metrologica, str. delle Cacce 91, 10135 Turin, Italy}
\affiliation{Istituto Nazionale di Fisica Nucleare, sez. di Torino, via P. Giuria 1, 10125 Turin, Italy}

%

\date{\today}

\begin{abstract}
    In this paper, we present a new experimental apparatus for the measurement of the detection efficiency of free-space single-photon detectors based on the substitution method. 
    For the first time, we extend the analysis to account for the wavelength dependence introduced by the transmissivity of the optical window in front of the detector's active area. 
    Our method involves measuring the detector's response at different wavelengths and comparing it to a calibrated reference detector. 
    This allows us to accurately quantify the efficiency variations due to the optical window's transmissivity. 
    The results provide a comprehensive understanding of the wavelength-dependent efficiency, which is crucial for optimizing the performance of single-photon detectors in various applications, including quantum communication and photonics research. 
    This characterization technique offers a significant advancement in the precision and reliability of single-photon detection efficiency measurements.
\end{abstract}

\maketitle

The detection of single photons is a critical component in a variety of scientific and technological applications, including quantum communication\cite{QComm1,QComm2,QComm3}, quantum computing\cite{QC1,QC2,QC3}, quantum imaging\cite{genovese2016real} and quantum sensing\cite{QSens1,QSens2,QSens3,petrini2020quantum} with photons. 
Accurate photon detection is essential for ensuring the reliability and precision of these applications. 
Consequently, the calibration of the detection efficiency of single-photon detectors (SPDs) is of paramount importance\cite{tara1,tara2,tara3,tara4,tara5,tara6,tara7,tara8,lopez2020study,tara10,tara11,tara12}.  
Such an efficiency is defined\cite{dictionary} as the probability of a SPD producing a measurable signal in response to one incident photon, depending on the wavelength and detection rate, with specific wavelength and count rate specifications. 

However, unlike detectors calibrated using classical radiometric techniques, there is currently no established standard for calibrating the detection efficiency of SPDs based on the measured counts. 
This lack of standardization presents a significant challenge, as it hinders the ability to ensure traceability of calibration results to the International System of Units (SI).
For this reason, a pilot study\cite{studioPilota} has recently been initiated among various national metrology institutes (NMIs) worldwide to attempt to define a characterization standard for free-space silicon single-photon avalanche diodes (Si-SPADs) detecting photons with a wavelength of $850\un{nm}$.
This collaborative effort aims to establish a unified and precise methodology for assessing the performance and detection efficiency of these detectors, thereby providing a reliable benchmark for scientific research and technological applications that depend on accurate single-photon detection.

Si-SPADs are SPDs operating in Geiger mode\cite{Tosi,SPAD1,SPAD2,SPDbook}. 
They are widely exploited due to their high detection efficiency in the visible range (up to $80\%$ for wavelengths around $650\un{nm}$), their low dark count rate (tens of counts per second), and their short dead time and jitter (typically tens of nanoseconds and hundreds of picoseconds, respectively). 
They can be exploited for a broad wavelength interval, from approximately $400\un{nm}$ to $1000\un{nm}$.
In particular, they find huge application for free-space Quantum Key Distribution (QKD)\cite{nielsen2001quantum}, wherein the employed wavelength is often around $850\un{nm}$, because of the transmissivity of the atmosphere just considering the range detectable from silicon-based SPDs, that is also the wavelength considered in the pilot study.

The detection efficiency is inherently dependent on the wavelength of the incident photons.
In addition, in SPDs operating in free-space with an optical window typically made of glass, there can be an additional (nonlinear) dependency on wavelength due to the interference effect that occurs between the two optical surfaces of the window, changing the transmissivity. 
Glass and quartz windows, while offering high transparency across a broad spectrum, still exhibit such a behavior, acting as an optical cavity, that can affect the overall detection efficiency.
These variations must be carefully characterized and accounted for to ensure accurate and reliable photon detection across different wavelengths. 
Furthermore, there is ongoing debate about interference effects due to a spatial nonuniform detector response, that could significantly impact the detection efficiency (see, e.g., the study carried on at NPL\cite{LukeChris}).

In this work, we present the INRiM new experimental setup for the measurement of the detection efficiency for free-space SPDs based on the substitution method\cite{lopez2015detection,lopez2020study}. 
In particular, we demonstrate such a technique on free-running Si-SPADs at $850\un{nm}$, in the framework of the aforementioned pilot study.
Additionally, we provide a model for the transmissivity of the quartz optical window to account for its impact on the overall detection efficiency.


\textit{Calibration technique.} 
The substitution method consists in a technique for comparing the signal measured by a SI-traceable detector with respect to the one measured from a SPD after a proper attenuation. 
This comparison regards very different light fluxes, deviating of several orders of magnitude. For example, a photon flux of about $1000\un{counts/s}$ with wavelength $850\un{nm}$ corresponds to an average optical power in the order of $10^{-16}W$.
Hence, the required attenuation between the high-flux regime and the single-photon level is usually around six or seven orders of magnitude, and it is of the utmost importance to characterize such an attenuation, containing its related uncertainty.

Then, the detection efficiency $\eta(\lambda)$ of a SPD  can be estimated comparing the macroscopic optical power $\mathcal{P}$ of a laser source, measured with a SI-traceable calibrated detector, and the rate $\mathcal{R}$ measured by the SPD after attenuating the same signal down to the single-photon level:
\begin{equation}\label{eq.model}
	\eta(\lambda)=\frac{hc}{\lambda}\frac{\mathcal{R}}{\tau \mathcal{P}}
\end{equation}
where $h$ is the Planck constant, $c$ is the speed of light in the vacuum and $\tau$ is the transmissivity due to the introduced attenuation. 

\begin{figure}[h!]
 \centering
	\includegraphics[width=1.1\columnwidth]{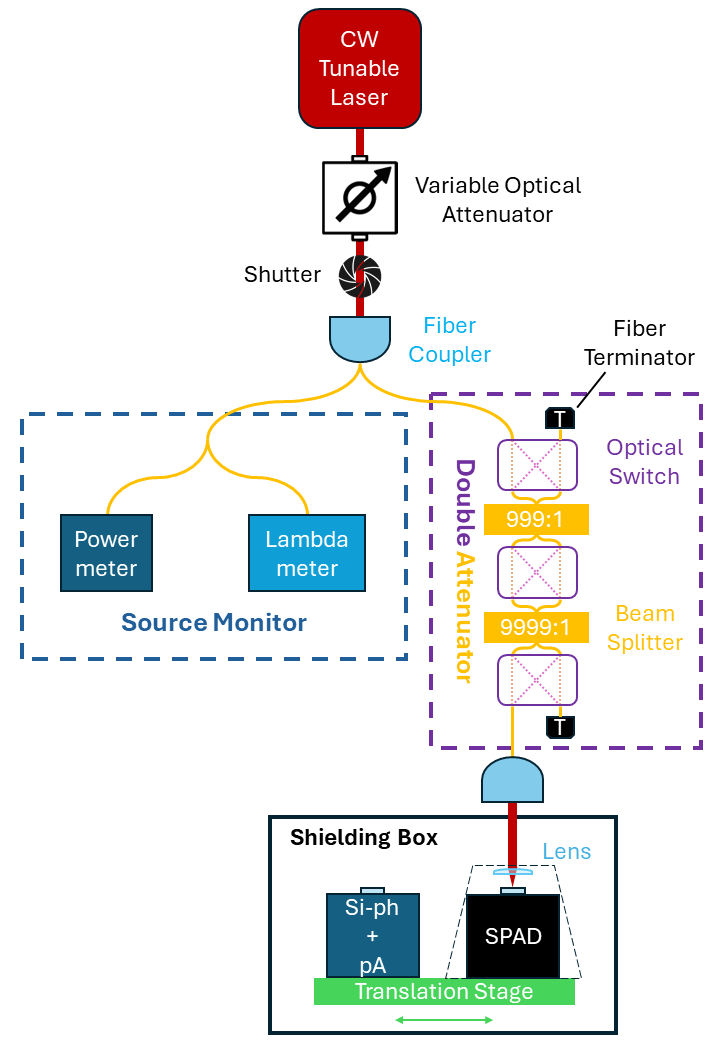}
	\caption{Experimental setup. The source is a tunable Ti:Sa laser followed by a variable optical attenuator.
    After a mechanical shutter, photons are fiber-coupled into a single-mode optical fiber through a $20\un{X}$ microscope objective, and two 50:50 beamsplitters address part of the photons to a powermeter and a lambdameter for monitoring the source stability. 
    A double attenuator system allows introducing the transmissivity $\tau$ addressing the photons through two unbalanced beam splitters: the former is a 999:1 ($30\un{dB}$), the latter 9999:1 ($40\un{dB}$). The selected path is controlled by three optical switches.
     Then, the output photons are directed through free space into a shielding box, where a motorized stage allows choosing between two detectors: a silicon photodiode (Si-ph) for high-flux measurements and the \textit{quantum} device under test (DUT) consisting in a free-space Si-SPAD in free-running mode. In front of the latter, photons are focused by a lens ($f = 8\un{mm}$). 
     The whole setup is fully automated and can be controlled through a LabVIEW programmed interface.}\label{fig.setup}
\end{figure}

In Fig. \ref{fig.setup} we show the measurement apparatus.
The source is a CW Ti:Sa laser with tunable wavelength.
The laser light intensity is controlled by a first variable optical attenuator stage exploiting polarizers and half waveplates. 
Then, after a mechanical shutter the laser light is fiber coupled into a single-mode fiber, optimizing the emitting spatial profile, and with two 50:50 beam splitters it is addressed to a monitor stage for checking both the emission wavelength and optical power stability, and to a second fully-pigtailed attenuator stage, that represents the novelty of our experimental apparatus. 
It consists of two unbalanced beam splitters: the former 999:1 and the latter 9999:1, respectively allowing introducing a $30\un{dB}$ and a $40\un{dB}$ attenuation. 
The light path is selected thanks to three  pc-controlled optical switches.
Selecting the path corresponding to the maximum attenuation (nominal $70\un{dB}$), the double-attenuator stage reproduces the transmissivity $\tau$ required in Eq. \ref{eq.model}.
Then, the photons will be out-coupled and collimated in free space with a Gaussian spatial mode to be sent into a shielding box for minimizing environmental photons.

A motorized translation stage allows selecting the measuring detector depending on the introduced attenuation: a SI-traceable Silicon photodiode (Si-ph) for the measurements of $\mathcal{P}$, and the Si-SPAD \textit{device under test} (DUT) for the measurement of $\mathcal{R}$, when the attenuation reproduces $\tau$ and the photon flux goes down to the single-photon level.

Since the usual diameter of a free-space Si-SPAD active area is about hundreds of micrometers, a lens focuses the photons spatial distribution on it in a diameter of $40\un{\mu m}$. The DUT position is optimized exploiting three actuators in the $x$,$y$ and $z$ directions.
All the devices of the depicted set-up are connected to a computer via LabVIEW interface, and the measurement routines are automatized. 

The DUT detection efficiency is estimated adapting Eq. \ref{eq.model} to the measurements realized with the experimental apparatus described in the previous section, i.e. comparing the SPAD measurements and the Si-ph output.
The SPAD produces a macroscopic pulse for each revealed photon and sends it to the electronics that collects $N$ counts in a time interval $t$, whereas the Si-ph generates photoelectrons proportionally to the incident optical power with a sensitivity $s$, and the resulting current is revealed by a picoamperometer with calibration factor $C$. 

The transmissivity $\tau$ is pc-controlled, and it has been independently measured (see below).
Furthermore, one has to consider also the transmissivity $T$ of the lens in front of the SPAD (see Fig. \ref{fig.setup}).
Finally, taking into account the (intrinsic and environmental) noise level for both the SPAD and the Si-ph measurements ($N^{\mathrm{env}}$ and $A^{\mathrm{env}}$), that account for dark counts and background photons as well as for the dark current. 
For the $i$-th measurement run Eq. \ref{eq.model} becomes:
\begin{equation}
    \label{eq.modsperimentale}
    \eta_i(\lambda)=\frac{h c}{\lambda t}\frac{s\left(N_i-N_i^{\mathrm{env}}\right)}{\tau C\left(A_i-A_i^{\mathrm{env}}\right)T}
\end{equation}

The transmittance $\tau$ is characterized exploiting the Si-ph detector. 
Our dual-attenuator approach enables us to divide the introduced attenuation, which is crucial for maintaining the linear response of the detector. 
Introducing the entire attenuation at once would compromise this linearity, increasing the related uncertainty and the accuracy and reliability of the experimental data.
Therefore,  $\tau$ is evaluated as  $\tau = \tau_{30\un{dB}}\tau_{40\un{dB}}$, where:
\begin{equation}
    \label{eq.tau}
    \tau_{x_{dB}} = \frac{A_{x\un{dB}} - A^{\mathrm{env}}}{A_{0\un{dB}} - A^{\mathrm{env}}}
\end{equation}
with the subscript $x$ indicating the selected attenuation by the optical switches (see Fig. \ref{fig.setup}).

\textit{\label{sec.results}Measurement and Results.} 
In our experiment, we characterize the behavior of a free-running Si-SPAD with a circular active area of $200\un{\mu m}$ in diameter with light at the wavelength of $\lambda = (850.711\pm 0.006)\un{nm}$.
Since the active area of a commercial free-space detector is generally not uniform, it is necessary to scan it to find a quite flat region.
\begin{figure}[h!]
	\centering
	\includegraphics[width=\columnwidth]{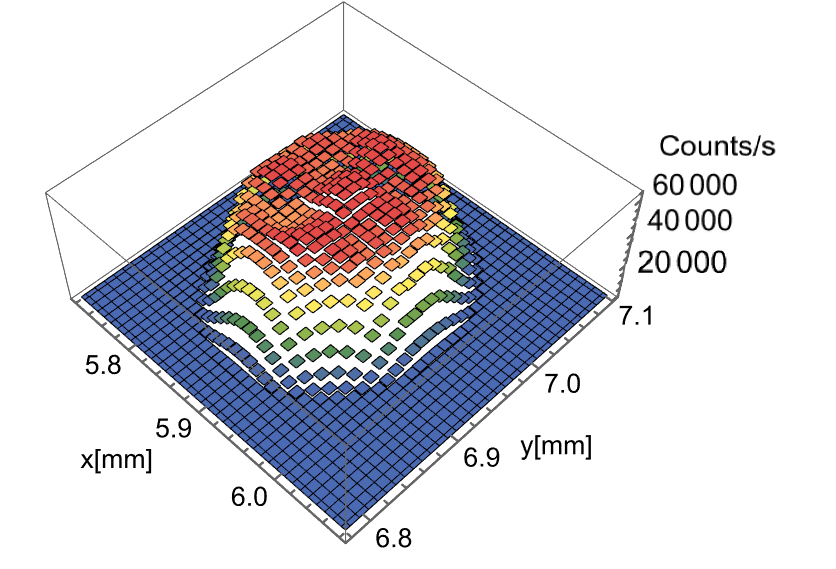}
	\caption{$(300\times 300)\un{\mu m^2}$ scansion of the Si-SPAD's active area obtained by setting a step size of $0.01\un{mm}$ in both the directions transversal to the photons propagation. 
    The distance with respect to the focal lens was previously optimized and fixed obtaining a focused beam with waist around $40\un{\mu m}$.}\label{fig.scan}
\end{figure}
Fig. \ref{fig.scan} shows that the surface of the active area is relatively uniform except for two dips located on the left side of the scan. 

Once the DUT position is fixed far from the two dips, we start the procedure for obtaining the DUT detection efficiency. 
First of all, we characterize $\tau$ (Eq. \ref{eq.tau}) averaging over a sequence of $100$ measurements.
Then, we repeat ten times the $\tau$ characterization in different days, i.e. evaluating the repeatability behavior of our double-attenuator system, obtaining $\tau = (2.1601\pm 0.0070)\times 10^{-7}$, highlighting a reasonable repeatability of our system day by day.

We underline that our double-attenuator approach allows good repeatability since it does not require to disconnect the optical fibers.
To monitor the source stability, we exploit the powermeter measurements at the source monitor stage (see Fig. \ref{fig.setup}).
This allows us to correct the measured $N_i$ and $A_i$ in Eq.s \ref{eq.modsperimentale},\ref{eq.tau} with respect to the source fluctuations according to:
\begin{equation}
    \begin{split}
        N_i &\rightarrow N'_i = N_i \rho_i^{\mathrm{DUT}}\\
        A_i &\rightarrow A'_i = \varepsilon A_i \rho_i^{\mathrm{Si-ph}}
    \end{split}
    \label{eq.correzione}
\end{equation}
where $\rho_i^{\mathrm{DUT}} = \langle P_i^{\mathrm{DUT}}\rangle/P_i^{\mathrm{DUT}}$ represents the correction with respect to the monitor powermeter measurement $P$ occurred during the $i$-th run of the DUT measurement $N_i$, meaning $\langle X_i \rangle$ the average value of the variable $X$; the same argument holds for $\rho_i^{\mathrm{Si-ph}}$.
Furthermore, since an unbalancement may happen between the average source optical power emission during the DUT and the Si-ph measurements, we have introduced the parameter $\varepsilon = \langle P_i^{\mathrm{DUT}}\rangle/\langle P_i^{\mathrm{Si-ph}}\rangle$.

To remove the arbitrary dependence on the count rate, we estimate the zero-flux efficiency $\eta_0$, i.e. the detection efficiency extrapolated to the zero-flux level, whose value is not affected by the presence of the SPD dead time.
The behavior of the detection efficiency can be described as\cite{Castelletto_2000,brida2000application,SPDbook}: 
\begin{equation}\label{eq.fit}
    \eta_{\lambda}(\langle N'_i\rangle) = \eta_0 -D \frac{\langle N'_i-N_{\mathrm{env}}\rangle}{t}
\end{equation}
where $D$ is the dead time and $N'_i$ is defined in Eq. \ref{eq.correzione}.
Hence, estimating $\eta_{\lambda}(\langle N'_i\rangle)$ for different photon fluxes, $\eta_0$ comes out from a linear regression.

\begin{figure}[h!]
	\centering
	\includegraphics[width=\columnwidth]{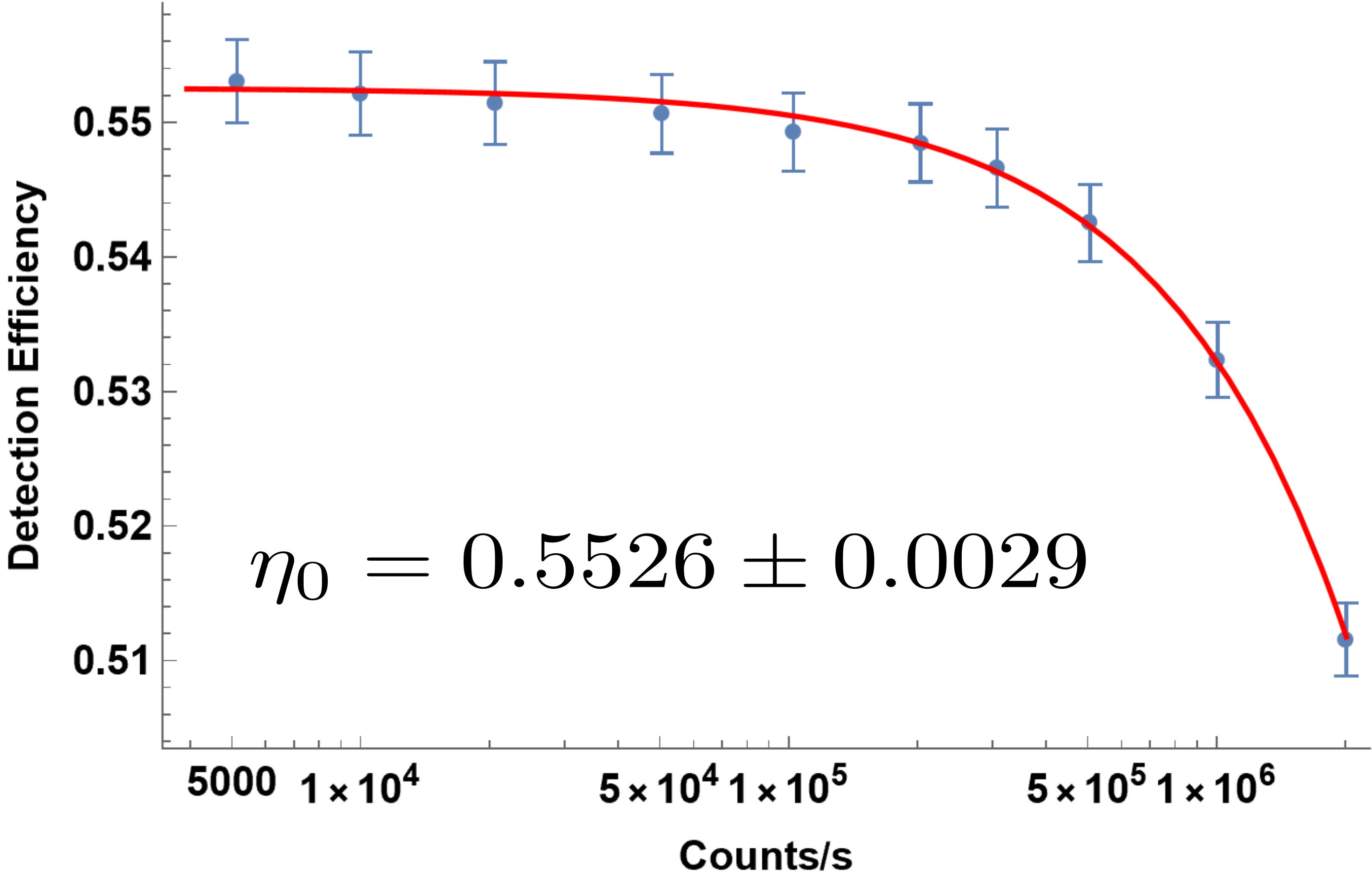}
	\caption{Detection efficiency estimation for different count rates $\mathcal{R} = \langle N_i' - N_i^{\mathrm{env}} \rangle / t$. The blue dots represent the calculated detection efficiencies with their related uncertainties, as determined by Eq. \ref{eq.modsperimentale}, while the red line is the result of the linear regression according to Eq. \ref{eq.fit}. $\eta_0$ is obtained as the intercept of the fit with the $y$-axis. All the shown uncertainties consider a coverage factor $K=1$.\label{fig.eff}}
\end{figure}

We measured the DUT detection efficiency (accordingly to Eq. \ref{eq.modsperimentale}) at different photon fluxes acting on the variable attenuator depicted in Fig. \ref{fig.setup}, obtaining various count-rate regimes from  $5000\un{counts/s}$ to $2\times 10^6\un{counts/s}$.
After collecting ten data points, we perform the linear regression as described in Eq. \ref{eq.fit} to estimate $\eta_0$. 
The results are shown in Fig. \ref{fig.eff}.

\begin{figure}[h!]
	\centering
	\includegraphics[width=\columnwidth]{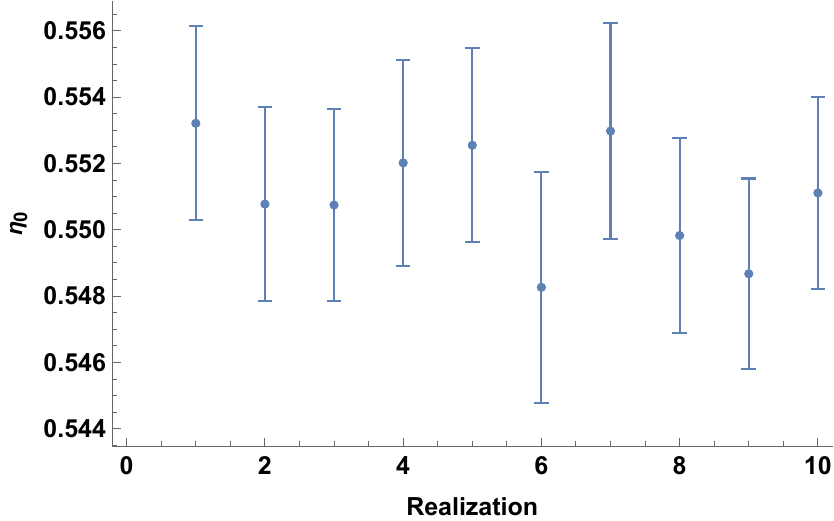}
	\caption{Si-SPAD detection efficiencies at zero-photon flux ($\eta_0$) obtained from ten independent estimations. The coverage factor of the uncertainties is $K=1$.}\label{fig.meas}
\end{figure}

As shown in Fig. \ref{fig.eff}, the experimental results align well with the fitted behavior. 
From the presented measurement run, we obtained $\eta_0 = 0.5526\pm 0.0029$ (coverage factor $K=1$). 
Uncertainties are propagated from Eq. \ref{eq.modsperimentale}, considering both statistical and non-statistical contributions.
An example of uncertainty budget for a fixed count rate is reported in Table \ref{tab}.
\begin{table}[h!]
\caption{Uncertainty budget related to the detection efficiency (see Eq. \ref{eq.modsperimentale}) $\langle \eta_i(N^{\prime},\lambda)\rangle$, with $N' = 20655\pm 27$, showed in Fig. \ref{fig.eff}.}
\begin{ruledtabular}
\label{tab}
\begin{tabular}{cccc}
Coefficient&Value&Uncertainty&$\%$ Contribution \\
\hline
$N'$ & $20655$ & $27$ & $5.47$\\
$N^{\mathrm{env}}$ & $28$ & $1$ & $0.012$ \\
$A'$ & $1.92807 \times 10^{-8}\un{A}$ & $4.9\times 10^{-12}\un{A}$ & $0.06$\\
$A^{\mathrm{env}}$ & $4.88 \times 10^{-14}\un{A}$ & $1.3\times 10^{-15}\un{A}$ & $1.5\times 10^{-8}$\\
$\tau$ & $2.1601 \times 10^{-7}$ & $7.0 \times 10^{-10}$ & $33.83$ \\
$\varepsilon$ & $1.0148$ & $1.4 \times 10^{-3}$ & $5.70$ \\
$s$ & $0.4766\un{W/A}$ & $1.9\times 10^{-3}\un{W/A}$ & $51.55$ \\
$C$ & $1.000023$ & $1.0 \times 10^{-5}$ & $3.2 \times 10^{-4}$ \\
$T$ & $0.985000$ & $3.0 \times 10^{-5}$ & $3.0 \times 10^{-3}$ \\
$\lambda$ & $8.50711\times 10^{-7}\un{m}$ & $6 \times 10^{-12}\un{m}$ & $1.6 \times 10^{-4}$ \\
$t$ & $1.0000\un{s}$ & $1.0 \times 10^{-3}\un{s}$ & $3.22$\\
\hline
$\langle\eta(N',\lambda)\rangle$ & $0.5514$ & $0.0031$
\end{tabular}
\end{ruledtabular}
\end{table}
Our approach demonstrates 
the possibility to measure the DUT detection efficiency in a SI-traceable manner, maintaining a contained uncertainty.
To assess the robustness of our technique, we repeat the entire estimation process ten times. 
The results are illustrated in Fig. \ref{fig.meas}.
Once more, the resulting estimations of $\eta_0$ exhibit strong agreement. 
The average value obtained for the detection efficiency at zero-photon flux is $\langle\eta_0\rangle = 0.5510\pm 0.0030$. 
This consistency across multiple estimations underlines the robustness and reliability of our measurement technique.

\begin{figure}[h!]
	\centering
	\includegraphics[width=\columnwidth]{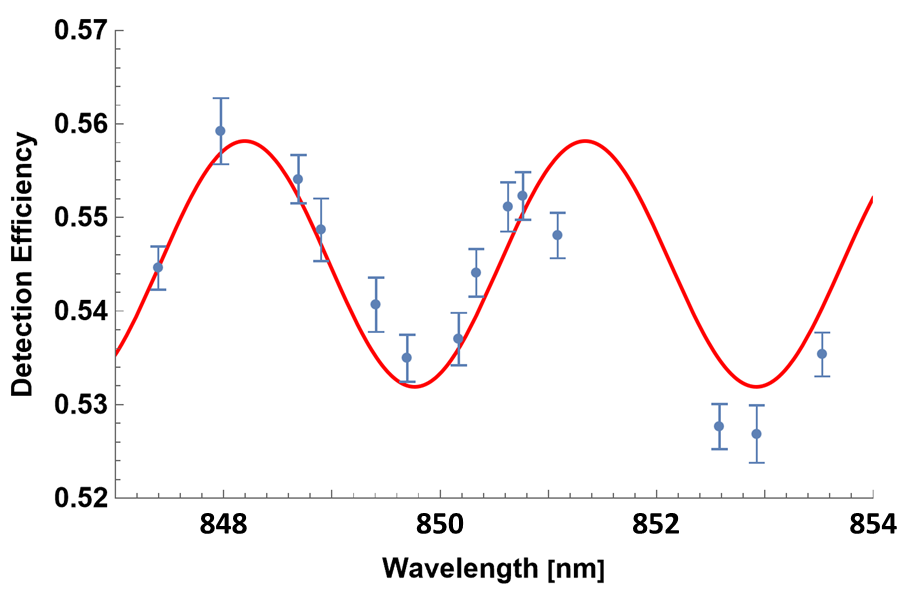}
	\caption{Characterization of detection efficiency as a function of wavelength. The number of detected photons for each acquisition lasting $t = 1\un{s}$ was fixed at approximately $N' \simeq 10^5$ counts. The blue dots represent the experimental results with their associated uncertainties ($K=1$), while the red line corresponds to the fitted model described by Eq. \ref{eq.genEff}.}\label{fig.lambda}
\end{figure}

Finally, we investigate the detection efficiency as a function of photon wavelength. 
To accomplish this, we replicate the estimation of $\langle\eta_i(\lambda)\rangle$ (Eq. \ref{eq.modsperimentale}) maintaining a constant $N'\simeq 10^5$, while varying the emission wavelength of our source. 
The value of $N'$ is arbitrarily chosen, and it represents a reasonable trade-off between reduced distortion effect due to SPD dead time and efficient data collection. 
The experimental data present a peculiar sinusoidal behavior (see Fig. \ref{fig.lambda}), that we interpret as an etaloning effect of the two surfaces of the optical window of the SPAD packaging.
Indeed, for a window with thickness $L$ and refractive index $n$, the transmissivity depends on the wavelength $\lambda$ through the parameter $\Gamma$, that is\cite{macleod2010thin}:
\begin{equation}
    \label{eq.Gamma}
    \Gamma(\lambda, n, L)=\frac{\gamma (1-\exp\left[{-2i\frac{2\pi}{\lambda} n L}\right])}{1-\gamma^2 \exp\left[{-2i\frac{2\pi}{\lambda} n L}\right]}
\end{equation}
where $\gamma = (n-n_a)/(n+n_a)$ and $n_a$ represents the air refractive index.
Then, the overall detection efficiency in Eq. \ref{eq.modsperimentale} can be generalized as:
\begin{equation}
    \label{eq.genEff}
    \eta_i(\lambda,n,L) = \eta_i(\lambda) \bigg(1-|\Gamma(\lambda,n,L)|^2\bigg)
\end{equation}
The results of this analysis are presented in Fig. \ref{fig.lambda}.

Our proposed model aligns closely with the experimental data. 
Moreover, these findings highlight the non-negligible impact of transmittance effects caused by the optical window. 
Within a range of approximately $3\un{nm}$, the detection efficiency varies by up to $5\%$. 
Consequently, it is imperative from a metrological perspective to consider such effects when characterizing free-space SPADs, rather than only focusing on a single wavelength.

\textit{Conclusions.} 
In this work, we presented the new experimental set-up for the measurement of the detection efficiency of free-space SPDs, exploiting the substitution method.
Specifically, the fully computer-controlled pigtailed attenuation stage optimizes the detection efficiency measurement time thanks to an excellent reproducibility, minimizing the uncertainty contribution associated to the attenuation measurement.
Then, we extended our analysis to the variation of the detection efficiency as a function of wavelength, taking into account the transmissivity of the optical window positioned in front of the detector's sensitive area. 
This comprehensive characterization is crucial for optimizing the performance of Si-SPADs in various applications, including quantum communication and photonics research. 
By understanding the wavelength dependence and the influence of the optical window, we can better estimate the efficiency of these detectors, leading to improved accuracy and reliability in single-photon detection. 
This study provides a valuable foundation for future metrological characterizations of Si-SPAD technology in both scientific and industrial contexts.
\hfill

This work was financially supported in the context of the following projects: Qu-Test project, which has received funding from the European Union’s Horizon Europe Research and Innovation Programme under grant agreement No. 101113901; QUID (QUantum Italy Deployment) and EQUO (European QUantum ecOsystems) projects which are funded by the European Commission in the Digital Europe Programme under the grant agreements number 101091408 and 101091561; the project G6026 SPS NATO; the EMPIR 19NRM06 METISQ and 23NRM04 NoQTeS. The projects 19NRM06 and 23NRM04 NoQTeS have received funding from the European Partnership on Metrology, co-financed from the European Union’s Horizon Europe Research and Innovation Programme and by the Participating States.

\hfill

\textit{Conflict of Interest.} 
The authors have no conflicts to disclose.

\textit{Author Contributions.}
All authors participated in the initial investigation on how to carry out the work under the supervision of IPD, AM, and GB. The experimental apparatus was built by SV and ER. Data acquisition was performed by ER under the supervision of SV and AM. Data analysis was conducted by SV, AM, and ER under the supervision of MGr and MGe. Data validation was carried out by AM and GB under the supervision of IPD. All authors contributed to the writing and revision of the article.
\hfill

\textit{Data Availability.} 
The data that support the findings of this study are available from the corresponding authors upon reasonable request.

\nocite{*}
\section*{References}
\bibliography{main}

\end{document}